\documentclass[preprint,12pt,authoryear]{elsarticle}

\usepackage{graphicx}
\usepackage{amsfonts}
\usepackage{amssymb}
\usepackage{amsmath}
\usepackage{amsthm}
\usepackage{url}
\usepackage{color}
\usepackage{hyperref}

\newtheorem{proposition}{Proposition}

\def\C{{\mathbf C}}
\def\F{{\mathbf F}}
\def\G{{\mathbf G}}
\def\H{{\mathbf H}}
\def\P{{\mathbf P}}
\def\Q{{\mathbf Q}}
\def\Z{{\mathbf Z}}
\def\I{{\mathbf I}}
\def\Om{{\mathbf \Omega}}
\def\M{{\mathbf M}}
\def\Tr{\mathrm{Tr}}
\def\T{\dagger}
\def\Gr{{\mathcal G}}

\begin{document}

\begin{frontmatter}

\title{Cleaning correlation matrices with external information: a review and the Maximum Variance estimator}

\author[mc,vr]{Sebastien Valeyre}
\ead{svaleyre@machinacap.com}
\ead{sv@valeyreresearch.com}
\address[mc]{Machina Capital, 21 rue de la Banque, 75002 Paris, France}
\address[vr]{Valeyre Research, 114 rue d'Antibes, 06400 Cannes, France}

\begin{abstract}
Rotationally invariant estimators clean the eigenvalues of an empirical
correlation matrix but keep its noisy eigenvectors. The Maximum Variance
(MaxVar) estimator corrects the eigenvectors with external information: the
leading eigenvectors are restricted to linear combinations of the
maximum-variance portfolios of predefined groups of assets, and the rest of the
spectrum is replaced by a single eigenvalue. The result is a parsimonious
factor model that remains positive definite when the observations are fewer
than the assets. We apply it to 47 futures on stock indices, bonds and
currencies, with the asset classes as groups. Over 1993--2018, the Sharpe ratio
of the risk parity portfolio is $1.28$ with the MaxVar estimator, against
$1.24$ with the rotationally invariant estimator and $1.17$ with the empirical
matrix, and the advantage increases for shorter estimation windows. We also
review the related approaches, from group averages and hierarchical principal
component analysis to eigenvector shrinkage, hierarchical clustering, and
projected and instrumented principal component analysis, and we show how the
estimator relates to each of them.
\end{abstract}


\end{frontmatter}

\section{Introduction}

The empirical correlation matrix of $n$ assets estimated from $T$ observations
is a noisy estimator of the true one as soon as $n$ is not small compared with
$T$. Random matrix theory describes this measurement noise: the small
eigenvalues are underestimated and the large ones overestimated
\citep{Marchenko67,Laloux99,Plerou99,Plerou02,Potters05}. The consequences for
portfolio construction are well known, since the mean--variance optimisation of
\citet{Markowitz52} involves the inverse of the matrix and therefore amplifies
the errors on the small eigenvalues \citep{Michaud89,Pafka03}.

Two families of remedies exist. The first one modifies the eigenvalues only and
keeps the empirical eigenvectors. It includes the eigenvalue clipping of
\citet{Laloux99}, the linear shrinkage towards the identity of
\citet{Ledoit04}, and the nonlinear shrinkage of \citet{Ledoit11,Ledoit12},
of which the quadratic-inverse shrinkage (QIS) of \citet{Ledoit22} is the most
recent form, and
whose implementation by \citet{Bun16,Bun17} is known as the rotationally
invariant estimator (RIE). The most recent developments keep this rotational
invariance and learn the transformation of the eigenvalues with a neural
network trained end to end to minimise the realised variance of the portfolio
\citep{Bongiorno25}. These estimators are optimal when nothing is known
about the eigenvectors, but they cannot correct them. The second family imposes
a structure on the matrix: shrinkage towards a single-index or
constant-correlation target \citep{Ledoit03,Ledoit04b}, block models
\citep{Engle12}, factor models with a sparse residual \citep{Fan13}, and
hierarchical models in which the structure is inferred from the data by
clustering \citep{Mantegna99,Tumminello07a,Tumminello07b,Bongiorno21}.
Here the eigenvectors are constrained, at the price of a bias when the assumed
structure is wrong. Hierarchical clustering has also been applied in the time
dimension, to covariance matrices estimated on different periods
\citep{Wang26}.

The use of predefined groups of assets to regularise correlations has a long
history. \citet{EltonGruber73} replaced the correlations by their averages
inside and between industries. More recently, \citet{Avellaneda20} introduced
a hierarchical principal component analysis (HPCA) in which the first
principal component of each predefined group is used as a factor: the
correlations between assets of different groups are reconstructed from the
correlations between these factors, while the empirical correlations inside
each group are kept. This construction is the closest relative of the
estimator studied here, and of the maximum-variance portfolios introduced
independently, shortly before, by \citet{Valeyre19}. In a different direction, \citet{Goldberg22},
\citet{Goldberg23} and \citet{Gurdogan22} correct the leading eigenvector
itself, by shrinking it towards directions that are known in advance, and
\citet{Goldberg24} adapt these directions to the portfolio that is optimised.
Finally, \citet{DeNard21} combine a factor model with a nonlinear shrinkage of
the covariance matrix of its residuals. A
related literature in asset pricing ties the factor loadings to observable
characteristics of the assets: the fundamental factor models initiated by
\citet{Rosenberg74}, the \textit{projected principal component analysis} of
\citet{Fan16} and the \textit{instrumented principal component analysis} of
\citet{Kelly19}. The relation of our estimator to these different approaches
is discussed in \ref{app:ipca} and \ref{app:shrinkhc}.

In practice, a portfolio manager does have information on the eigenvectors:
the assets belong to classes, regions or sectors that are known in advance and
do not depend on the sample. In earlier work on single stocks,
\citet{Valeyre19} used such information to clean the correlation matrix: for
each financial theme (book value, capitalisation, sector, etc.) they built the
market-neutral portfolio of maximum variance, which amounts to taking the
second eigenvector of the theme, after that of the market, and
they diagonalised the correlation matrix in the subspace spanned by these
``Fundamental Maximum Variance Portfolios''. The resulting constrained
eigenvectors are linear combinations of economically meaningful portfolios, in
the same way as molecular orbitals are approximated by linear combinations of
atomic orbitals in quantum chemistry. They also found that the sensitivity
of a stock to the factor associated with a characteristic is a linear function
of the rank of the stock according to this characteristic, and that the same
function describes most characteristics. Weights that are linear in the rank
are commonly used to build factor portfolios, as a convenient way of limiting
the influence of outliers, but, to our knowledge, this form had never been
measured. This result and a possible interpretation are presented in
\ref{app:rank}.

The purpose of this paper is to illustrate this chapter of the doctoral thesis
of the author beyond the world of single stocks, in which it was developed. The
thesis seems to have remained largely unread, and a second purpose is to make
its results more accessible. We apply the same idea to a
cross-asset universe of futures on stock indices, bonds and currencies, where
the financial characteristics of stocks are replaced by the asset classes. To
this end, we formulate the idea as a general cleaning method for a correlation
matrix, we establish its properties, and we compare it with the rotationally
invariant estimator. Our contributions are the following.
\begin{enumerate}
\item We define the Maximum Variance (MaxVar) estimator for an arbitrary family
of groups of assets, possibly overlapping, and show that it is a factor model
that is positive definite and does not depend on any arbitrary choice
(Section~\ref{sec:theory}). We show that its central step, a principal
component analysis under constraint, is equivalent to the \textit{projected
principal component analysis} of \citet{Fan16}, with the difference that the
portfolios that span the projection
subspace are themselves estimated from the data, and we relate it to the factor
models used in asset pricing
(\ref{app:ipca}). The diagonalisation of the correlation matrix in the
subspace of predefined maximum-variance portfolios was introduced by
\citet{Valeyre19}; the use of the first principal component of predefined
groups was proposed independently in the HPCA of \citet{Avellaneda20}, with
which our estimator coincides for the correlations between groups. Our contribution here lies in the
treatment of overlapping groups and in the regularisation of the correlations
inside the groups, which yields a parsimonious matrix that remains invertible
when the observations are fewer than the assets (\ref{app:shrinkhc}).
\item We illustrate the reduction of the measurement noise by a Monte Carlo
simulation in which the reference matrix has the structure assumed by the
estimator (Section~\ref{sec:mc}).
\item We apply the MaxVar estimator to 47 futures on stock indices, bonds and
currencies and compare the Sharpe ratios of the risk parity portfolio built with
the empirical matrix, with the rotationally invariant estimator and with the
MaxVar estimator
(Section~\ref{sec:backtest}).
\end{enumerate}

\section{The Maximum Variance estimator}
\label{sec:theory}

\subsection{Definition}

Let $\C$ be the $n\times n$ empirical correlation matrix of $n$ assets. The
external information consists of $K<n$ groups of assets $\Gr_1,\ldots,\Gr_K
\subset\{1,\ldots,n\}$, whose union contains all the assets. The groups may
overlap: for futures, one can use asset classes (stock indices, bonds,
currencies) together with regions (Europe, America, Asia). The estimator is
built in four steps.

\paragraph{Step 1: maximum-variance portfolio of each group} For each group
$\Gr_k$, let ${\bf p}_k$ be the portfolio of unit norm, invested in the assets
of $\Gr_k$ only, that has the largest variance:
\begin{equation}
\label{eq:pk}
{\bf p}_k = \arg\max\limits_{{\bf w}}\ {\bf w}^\T \C\, {\bf w} \qquad
\text{subject to}\quad {\bf w}^\T{\bf w}=1, \quad w_i=0 \ \text{ for } i\notin\Gr_k .
\end{equation}
It is the leading eigenvector of the restriction of $\C$ to the group,
completed by zeros. We denote by $\P=({\bf p}_1,\ldots,{\bf p}_K)$ the
$n\times K$ matrix of these portfolios, which are assumed to be linearly
independent.

\paragraph{Step 2: constrained eigenvectors} We look for the eigenvectors of
$\C$ under the constraint that they are linear combinations of the ${\bf
p}_k$, i.e., for the stationary points of ${\bf w}^\T\C{\bf w}/{\bf w}^\T{\bf
w}$ with ${\bf w}=\P{\bf a}$. With
\begin{equation}
\G = \P^\T\P , \qquad \H = \P^\T\C\,\P ,
\end{equation}
this is the generalised eigenvalue problem $\H{\bf a}=\mu\,\G{\bf a}$. The
matrix $\G$ is the identity when the groups are disjoint. Diagonalising the
symmetric $K\times K$ matrix
\begin{equation}
\G^{-1/2}\,\H\,\G^{-1/2} = \Om\,\M\,\Om^\T , \qquad \M=\mathrm{diag}(\mu_1,\ldots,\mu_K) ,
\end{equation}
we obtain the constrained eigenvalues $\mu_1\geq\ldots\geq\mu_K$ and the
constrained eigenvectors, which are the columns of
\begin{equation}
\label{eq:Z}
\Z = \P\,\G^{-1/2}\,\Om .
\end{equation}
They are orthonormal and they diagonalise $\C$ in the subspace: $\Z^\T\Z=\I_K$
and $\Z^\T\C\,\Z=\M$. This principal component analysis under constraint is
equivalent to the \textit{projected principal component analysis} of
\citet{Fan16}: the constrained
eigenvalues and eigenvectors are those of the correlation matrix of the
returns projected onto the subspace of the group portfolios. The equivalence,
and the relation to the projected and instrumented principal component analyses
of the asset-pricing literature, are detailed in \ref{app:ipca}.

\paragraph{Step 3: completion} In the orthogonal complement of the subspace
spanned by the ${\bf p}_k$, no information is available, and we replace $\C$ by
a multiple of the identity that preserves the trace:
\begin{equation}
\label{eq:F}
\F = \Z\,\M\,\Z^\T + \delta\,\bigl(\I_n-\Z\Z^\T\bigr) , \qquad
\delta = \frac{n-\sum_{k=1}^K\mu_k}{n-K} .
\end{equation}

\paragraph{Step 4: normalisation} The diagonal elements of $\F$ are
\begin{equation}
\label{eq:diag}
\F_{ii} = \delta + \sum\limits_{k=1}^K (\mu_k-\delta)\, z_{k,i}^2 ,
\end{equation}
where $z_{k,i}$ is the weight of asset $i$ in the $k$-th constrained
eigenvector. They are equal to $1$ on average, since $\Tr\,\F=n$, but not
individually: $\F_{ii}$ is smaller than $1$ for an asset that is weakly exposed
to the constrained eigenvectors and larger than $1$ for a strongly exposed one.
They are all equal to $1$ only if the assets of each group have the same
exposure. The MaxVar estimator is therefore defined as the correlation matrix
associated with $\F$,
\begin{equation}
\label{eq:Fhat}
\hat\F_{ij} = \frac{\F_{ij}}{\sqrt{\F_{ii}\,\F_{jj}}} .
\end{equation}
This normalisation has a simple interpretation. Writing
$\F=\delta\,\I_n+\Z(\M-\delta\,\I_K)\Z^\T$, the normalised matrix $\hat\F$ is a
standard $K$-factor model when $\mu_K\geq\delta$, which is the case here: the
loadings of asset $i$ are those of $\F$ divided by $\sqrt{\F_{ii}}$, and its
specific variance is $\delta/\F_{ii}$. An asset
that is weakly exposed to the factors thus has a specific variance close to
$1$, as it should. The constrained eigenvectors are eigenvectors of $\F$ and
not of $\hat\F$: the normalisation modifies them. This is the usual situation
in factor models, whose loadings are not eigenvectors of the covariance matrix
when the specific variances differ across assets.

An alternative to the normalisation of Eq.~(\ref{eq:Fhat}), which preserves
the constrained eigenvectors exactly, is to
replace the common residual variance $\delta$ of Eq.~(\ref{eq:F}) by a residual
variance specific to each asset,
\begin{equation}
	\label{eqalt}
	\tilde\F = \Z\,\M\,\Z^\T + \Q\,{\bf \Delta}\,\Q^\T , \qquad \Q=\I_n-\Z\Z^\T ,
\end{equation}
where ${\bf \Delta}=\mathrm{diag}(\Delta_1,\ldots,\Delta_n)$ is the solution of
the linear system
\begin{equation}
	\label{eqlinearsyst}
	\sum_{j=1}^n \left(\Q_{ij}\right)^2\Delta_j = 1-\sum_{k=1}^K \mu_k\, z_{k,i}^2 ,
	\qquad i=1,\ldots,n .
\end{equation}
The diagonal elements of $\tilde\F$ are then exactly equal to $1$, and the
constrained eigenvectors remain eigenvectors of $\tilde\F$. Eq.~(\ref{eqalt})
reduces to Eq.~(\ref{eq:F}) when all the $\Delta_i$ are equal. The
choice between the two is a trade-off between the exactness of the
eigenvectors and the conditioning of the matrix: $\hat\F$ is always positive
definite, with eigenvalues bounded from below by
$\min(\delta,\mu_K)/\max_i\F_{ii}$,
whereas $\tilde\F$ is positive definite only if all the $\Delta_i$ are
positive, and its smallest eigenvalues can be much smaller than $\delta$. The
difference between the two is small for the portfolio considered here, with an
impact of at most $0.05$ on the Sharpe ratio, and we use $\hat\F$ in the rest
of the paper.

\subsection{Properties}

\begin{proposition}
\label{prop:prop}
{\rm (i)} $\F$ depends on the groups only through the subspace spanned by the
${\bf p}_k$, and it does not depend on the choice of a basis of the orthogonal
complement. {\rm (ii)} $\Tr\,\F=n$ and $\delta=\Tr(\Q\C\Q)/(n-K)\geq 0$, where
$\Q=\I_n-\Z\Z^\T$. {\rm (iii)} If $\C$ is positive definite on the subspace
spanned by the ${\bf p}_k$ and does not vanish on its complement, then $\F$ and
$\hat\F$ are positive definite, even if $\C$ is singular. {\rm (iv)} The
constrained eigenvalues are bounded by the eigenvalues
$\lambda_1\geq\ldots\geq\lambda_n$ of $\C$: $\lambda_{n-K+k}\leq\mu_k\leq
\lambda_k$.
\end{proposition}

\begin{proof}
(i) The matrix $\Z\M\Z^\T$ is the compression of $\C$ to the subspace, which
does not depend on the basis, and $\Q$ is the projector onto the complement.
(ii) $\Tr\,\F=\sum_k\mu_k+\delta(n-K)=n$, and $\Tr(\Q\C\Q)=\Tr\,\C-\Tr(\Z^\T
\C\Z)=n-\sum_k\mu_k$; it is non-negative because $\Q\C\Q$ is positive
semidefinite. (iii) Under the assumptions, $\mu_k>0$ and $\delta>0$, and the
eigenvalues of $\F$ are the $\mu_k$ and $\delta$ (with multiplicity $n-K$); the
normalisation (\ref{eq:Fhat}) is a congruence and preserves positivity. (iv)
This is the Poincar\'e separation theorem applied to $\Z^\T\C\Z$.
\end{proof}

Property (iii) is of practical importance: the inverse of $\hat\F$, which is
needed for the portfolio of Section~\ref{sec:backtest}, exists even when the
number of observations is smaller than $n$.

\paragraph{Disjoint groups} When the groups are disjoint, the ${\bf p}_k$ are
orthonormal and the structure of $\F$ is transparent. Let $\lambda_k^{(0)}=
\H_{kk}$ be the variance of ${\bf p}_k$. For two different assets $i\in\Gr_k$
and $j\in\Gr_l$,
\begin{equation}
\label{eq:disjoint}
\F_{ij} =
\begin{cases}
\bigl(\lambda_k^{(0)}-\delta\bigr)\, p_{k,i}\, p_{k,j} & (k=l), \\[1mm]
\H_{kl}\, p_{k,i}\, p_{l,j} & (k\ne l) .
\end{cases}
\end{equation}
Inside each group, the correlations are those of a one-factor model whose
loadings are given by the maximum-variance portfolio of the group; between two
groups, they are the product of the loadings by the covariance $\H_{kl}$ of the
two group portfolios. The estimator has $n-K$ free loadings and $K(K+1)/2$
elements of $\H$, i.e., about $n+K^2/2$ parameters instead of $n(n-1)/2$
correlations: for the 47 futures of Section~\ref{sec:backtest} and three asset
classes, 50 parameters instead of 1081.

\paragraph{Limiting cases} With a single group that contains all the assets,
${\bf p}_1$ is the leading eigenvector of $\C$ and $\F$ is the matrix obtained
by clipping all the eigenvalues but the largest one. In the opposite limit of groups
reduced to single assets, the subspace is the whole space, $\F=\C$ and nothing
is cleaned. The family of groups thus
interpolates between a one-factor model and the raw empirical matrix.

\paragraph{Rotational invariance} Unlike the RIE, the MaxVar estimator is not
rotationally invariant: the groups single out directions in the space of
assets. This is the purpose of the method, and its risk. If the groups are
informative, the eigenvectors are improved, which no rotationally invariant
estimator can achieve. If the true matrix has a structure that the groups do
not describe, this structure is erased.

\subsection{Bias and variance}
\label{sec:bias}

Applied to the true correlation matrix $\C_0$, the MaxVar map does not return
$\C_0$ unless $\C_0$ has exactly the structure (\ref{eq:F}). The estimator has
therefore a bias that does not depend on the number of observations $T$, in
addition to a variance that is small because few parameters are estimated. By
contrast, the empirical matrix has no bias and a variance of order $n^2/T$. The
MaxVar estimator is thus expected to be most useful for short samples, and for
applications that are more sensitive to the measurement noise than to the fine
structure of the correlations, such as portfolios that involve the inverse of
the matrix. Whether the reduction of the noise outweighs the bias for a given
universe and a given use of the matrix is an empirical question, which we
address in Section~\ref{sec:backtest}.

\section{Monte Carlo illustration}
\label{sec:mc}

We first illustrate the reduction of the measurement noise on a simulated
example. The reference matrix $\C_0$ is the correlation matrix of the 47
futures of Section~\ref{sec:backtest} measured on 28 December 2018 and cleaned
by the MaxVar estimator with the three asset classes as groups
(Fig.~\ref{fig:map}(a)). We generate 100 samples of $T=150$ independent Gaussian
vectors of covariance $\C_0$, which corresponds to the 150 weekly returns of the
estimation window used in Section~\ref{sec:backtest}, and compute for each
sample the empirical correlation matrix $\C$. This noisy matrix is then cleaned
by the MaxVar estimator and by the rotationally invariant estimator (RIE) of
\citet{Bun16,Bun17}, and each estimate $\hat\C$ is compared with the reference
through the Frobenius distance
\begin{equation}
\label{eq:distance}
d(\hat\C,\C_0) = \sqrt{\Tr\bigl[(\hat\C-\C_0)^\T(\hat\C-\C_0)\bigr]} .
\end{equation}

\begin{table}[t]
\centering
\begin{tabular}{l c}
\hline
Estimator & Distance to the reference \\
\hline
MaxVar    & 2.09 \\
RIE       & 2.99 \\
Empirical & 3.24 \\
\hline
\end{tabular}
\caption{Average Frobenius distance (\ref{eq:distance}) between the reference
correlation matrix and its estimates from 150 simulated weekly returns of the
47 futures (100 samples). The reference is the MaxVar matrix of 28 December
2018.}
\label{tab:mc}
\end{table}

The results are given in Table~\ref{tab:mc}. The MaxVar estimator reduces the
distance to the reference by one third with respect to the empirical matrix
($2.09$ against $3.24$), whereas the RIE reduces it by less than $10\%$
($2.99$).

This test must be read with caution. The reference matrix has, by construction,
the structure assumed by the MaxVar estimator, so that the specification is
favourable to it: the test measures how much of the measurement noise each
method removes when the assumed structure is correct, and it says nothing about
the bias discussed in Section~\ref{sec:bias}. It is therefore an illustration
of the potential of the method rather than a proof of its superiority. The
measure that matters in the end is the performance of a portfolio built with the
cleaned matrices on real data, to which we now turn.

\section{Application to futures}
\label{sec:backtest}

\subsection{Data and portfolio}

We consider the most liquid futures: 24 on stock indices, 14 on government
bonds and 9 on currencies (Table~\ref{tab:listofinstrument}). The data start on
8 May 1985 and end on 31 December 2018; the statistics are computed over the
period from 1 January 1993 to 31 December 2018. Transaction costs and market
impact are not taken into account.

The daily returns $r_t^j$ are rescaled by the volatility $v_t^j$ of each
contract, estimated by an exponential moving average (EMA) of the squared
daily returns with a parameter $\eta=1/100$,
\begin{equation}
[v_t^j]^2 = (1-\eta)\, [v_{t-1}^j]^2 + \eta\, [r_t^j]^2 .
\end{equation}
The correlation matrix is estimated from weekly returns $\hat r_t^j$, to
offset the different trading hours of the markets, with an EMA of parameter
$\eta'$,
\begin{equation}
\hat\C_t^{ij} = (1-\eta')\, \hat\C_{t-1}^{ij} + \eta'\, \hat r_t^i\, \hat r_t^j ,
\qquad
\C_t^{ij} = \frac{\hat\C_t^{ij}}{\sqrt{\hat\C_t^{ii}\,\hat\C_t^{jj}}} .
\end{equation}
With $\eta'=1/750$ in units of days, the estimation window contains about 150
independent weekly returns for 47 contracts. The empirical matrix
$\C_t$ is then cleaned either by the RIE or by the MaxVar estimator with the
three asset classes as groups ($K=3$).

We compare the cleaned matrices through the Sharpe ratio of the risk parity
portfolio,
\begin{equation}
\label{eq:Pi_RP}
\Pi_t \propto \Sigma_t^{-1}\, \C_t^{-1}\, {\bf e} ,
\end{equation}
where $\Sigma_t$ is the diagonal matrix of the volatilities and ${\bf e}$ is
the vector whose components are equal to $1$ for the stock indices and the
bonds and to $0$ for the currencies. This portfolio follows the logic of the
maximum diversification portfolio of \citet{Choueifaty08}: it is optimal if
all the stock indices and bonds have the same Sharpe ratio and if the
currencies have no expected return. The currencies are nevertheless held,
through their correlations with the other contracts, as hedges of the
positions in stock indices and bonds. This portfolio depends on the
data only through the volatilities and the correlation matrix, so that its
performance isolates the effect of the cleaning method; other uses of the
correlation matrix are not considered here.

\begin{table}[t]
\centering
\small
\begin{tabular}{p{5.2cm} p{4.0cm} p{3.0cm}}
\hline
Stock indices & Bonds & Currencies \\
\hline
\raggedright AEX, S\&P Midcap 400, Russell 2000, CAC 40, DAX, FTSE 100, STOXX
Europe 600, Hang Seng, FTSE MIB, Nikkei 225, Topix, Kospi 200, MSCI EAFE, IBEX
35, MSCI Emerging Markets, Nifty, Nasdaq 100, MSCI Singapore, S\&P 500, MSCI
Taiwan, Euro Stoxx 50, S\&P Canada 60, SPI 200, Dow Jones &
\raggedright Bund, Canada 10Y, Bobl, Schatz, BTP, Buxl, Gilt, OAT, US T-note
5Y, JGB, US T-note 10Y, US T-note 2Y, US Ultra T-bond, US T-bond 30Y &
\raggedright AUD, GBP, CAD, EUR, US dollar index, JPY, MXN, NZD, CHF
\tabularnewline
\hline
\end{tabular}
\caption{List of the 47 futures, in the order used in Fig.~\ref{fig:map}.
Currencies are quoted against the US dollar.}
\label{tab:listofinstrument}
\end{table}

\subsection{Cleaned matrices}

\begin{figure}[t]
\begin{center}
\includegraphics[width=0.32\textwidth]{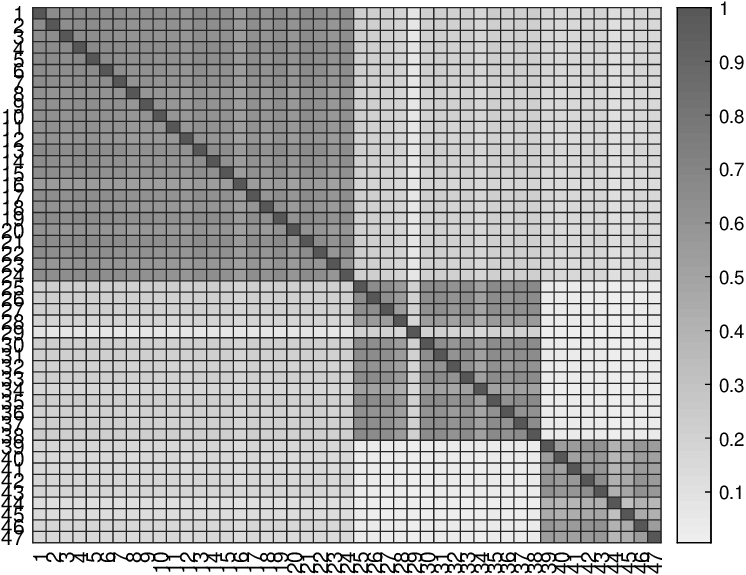}\hfill
\includegraphics[width=0.32\textwidth]{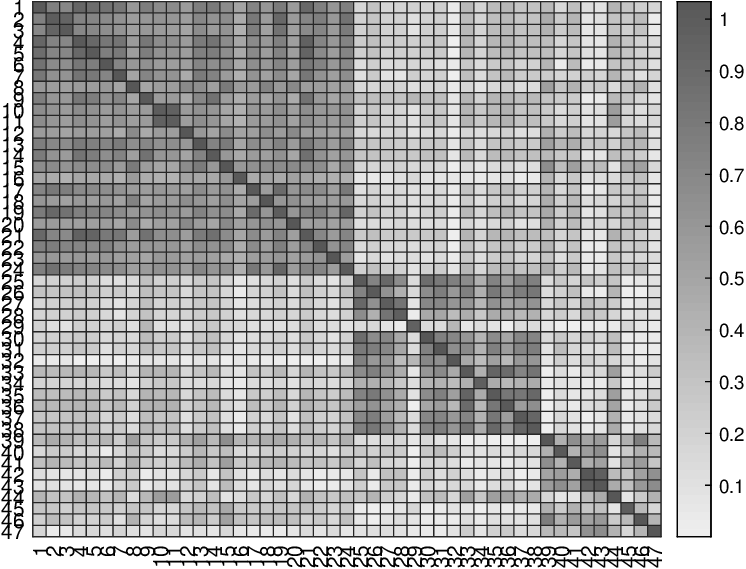}\hfill
\includegraphics[width=0.32\textwidth]{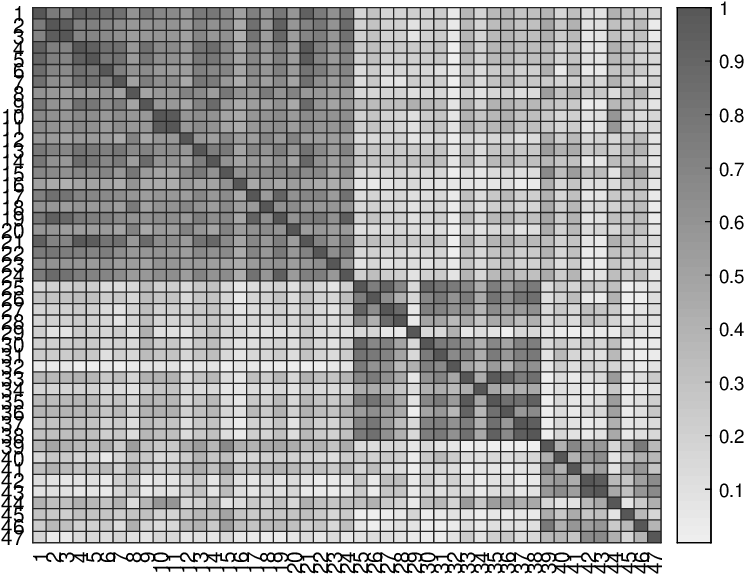}
\end{center}
\caption{Heat maps of the absolute values of the correlations between the 47
futures on 28 December 2018 ($\eta'=1/750$). Left: MaxVar estimator with the
three asset classes as groups. Middle: RIE. Right: empirical matrix. The
contracts are ordered as in Table~\ref{tab:listofinstrument}: stock indices
(1--24), bonds (25--38) and currencies (39--47).}
\label{fig:map}
\end{figure}

Figure~\ref{fig:map} shows the absolute values of the correlations measured on
28 December 2018, before and after cleaning. The empirical matrix has a clear
block structure. Some contracts are less correlated with their class than the
others: among the stock indices, the Nifty (16th position; $0.55$ against
$0.60$); among the bonds, the Italian BTP (5th bond; $0.20$ against $0.60$);
among the currencies, the Mexican peso and the yen.

The RIE hardly changes the appearance of the matrix, since it modifies the
eigenvalues only. The MaxVar estimator has a stronger effect. It keeps the
block structure and the specific behaviour of the weakly correlated contracts,
through their small loadings, and it smooths the correlations inside each
block, in agreement with Eq.~(\ref{eq:disjoint}). Part of this smoothing removes
measurement noise; part of it may remove genuine differences between contracts
of the same class, which is the bias discussed in Section~\ref{sec:bias}. The
heat maps alone cannot tell the two apart, which is why we rely on the
performance of the portfolio.

\subsection{Performance of the risk parity portfolio}

\begin{table}[t]
\centering
\begin{tabular}{l c c c}
\hline
Estimation window & Empirical & RIE & MaxVar \\
\hline
$\eta'=1/750$      & 1.17 & 1.24 & 1.28 \\
$\eta'=1/100$      & --   & 0.93 & 1.05 \\
\hline
\end{tabular}
\caption{Sharpe ratio over 1993--2018, before costs, of the risk parity
portfolio (\ref{eq:Pi_RP}) built with the empirical correlation matrix and with
the matrices cleaned by the RIE and by the MaxVar estimator ($K=3$), for two
values of the parameter $\eta'$ of the estimation window.}
\label{tab:sharpe}
\end{table}

Table~\ref{tab:sharpe} gives the Sharpe ratios. Cleaning the matrix improves
the performance of the risk parity portfolio, which involves the inverse
matrix: the Sharpe ratio increases from $1.17$ with the empirical matrix to
$1.24$ with the RIE and $1.28$ with the MaxVar estimator. When the estimation
window is shortened to $\eta'=1/100$, i.e., to about 20 weekly returns for 47
contracts, the advantage of the MaxVar estimator over the RIE increases
($1.05$ against $0.93$), as expected for an estimator of small variance.

A caveat is in order. The differences between the Sharpe ratios of
Table~\ref{tab:sharpe} are small compared with the statistical error of a
Sharpe ratio measured over 26 years, which is of the order of $1/\sqrt{26}
\approx 0.2$. Since the portfolios built with the three matrices are strongly
correlated, the differences are measured more precisely than each Sharpe ratio,
but their statistical significance has not been tested. Moreover, these results
concern a single portfolio, and they do not establish that the estimator is
preferable for other uses of the correlation matrix.

\section{Conclusion}
\label{sec:conclusion}

This paper illustrates, in a cross-asset universe, an approach that was
introduced for single stocks by \citet{Valeyre19}. We have formulated the
Maximum Variance estimator of a correlation matrix, which uses a predefined
family of groups of assets to constrain the leading eigenvectors, and applied
it to futures on stock indices, bonds and currencies, with the asset classes
in the role played by the financial characteristics for stocks. The estimator is a parsimonious factor model, always positive
definite, with about $n$ parameters. Unlike the rotationally invariant
estimator, it corrects the eigenvectors and not only the eigenvalues, at the
price of a bias when the true matrix departs from the assumed structure. It
belongs to the family of projected and instrumented principal component
analyses, from which it differs in two ways: the weights of the assets inside
each group are estimated from the data instead of being imposed, and the
factor part is completed into a full correlation matrix that can be inverted.

In a Monte Carlo test whose reference matrix has the structure assumed by the
estimator, a specification that is favourable to it and that we use as an
illustration, the MaxVar estimator removes a much larger part of the
measurement noise than the rotationally invariant estimator. The test that
matters is the performance on real data. On 47 futures over 1993--2018, the
MaxVar estimator with the three asset classes as groups gives the highest
Sharpe ratio for the risk parity portfolio, which involves the inverse of the
matrix, and its advantage over the rotationally invariant estimator increases
when the estimation window is shortened.

Several extensions are natural: testing the statistical significance of the
differences between Sharpe ratios, taking the groups from a hierarchical
clustering of the assets rather than from an external classification, assessing the estimator on portfolios that
use the correlation matrix in other ways, using groups finer than the asset
classes
to reduce the bias, and using the MaxVar matrix as a shrinkage target in the
spirit of \citet{Ledoit03}, so as to combine its small variance with the
absence of bias of the empirical matrix. A direct comparison, on the same
data, of the three treatments of the correlations inside the groups (empirical
as in \citealp{Avellaneda20}, shrunk as in \citealp{DeNard21}, or replaced by
a flat spectrum as here) as a function of the number of observations per asset
would measure the trade-off between bias and variance that is at the heart of
the method.

Finally, this paper recalls a result of \citet{Valeyre19} that deserves
attention in its own right: for most characteristics of stocks, the
sensitivity to the associated factor is proportional to the centred rank of the
stock according to this characteristic (\ref{app:rank}). Rank-based weights
are used routinely to build factor portfolios, as a convention. It is
therefore surprising that, seven years later, this measurement has, to our
knowledge, been neither taken up nor tested independently. We hope that the
interpretation proposed here, and the prediction that it makes about the
extreme quantiles, will encourage such tests.

\appendix
\section{Relation to projected and instrumented principal component analysis}
\label{app:ipca}

The second step of the MaxVar estimator has a close counterpart in the
asset-pricing literature on factor models whose loadings are tied to observable
characteristics of the assets. This appendix makes the link explicit and states
what is specific to the present approach.

\subsection{Constrained diagonalisation and the projected principal component analysis of Fan et al.}

Let ${\bf r}_t$ ($t=1,\ldots,T$) be the vectors of normalised returns, so that
$\C=T^{-1}\sum_t{\bf r}_t{\bf r}_t^\T$, and consider the factor model
\begin{equation}
\label{eq:ipca_model}
{\bf r}_t = \P\,{\bf \Gamma}\,{\bf f}_t + {\bf e}_t ,
\end{equation}
in which the $n\times K$ matrix $\P$ is given, ${\bf \Gamma}$ is a $K\times K'$
matrix of coefficients with $K'\leq K$, and ${\bf f}_t$ is a vector of $K'$
latent factors. The loadings $\P{\bf \Gamma}$ of the assets are thus
constrained to be linear combinations of the columns of $\P$. Each column is a
vector of weights on the $n$ assets, and we call these columns the basis
portfolios. In the econometric literature they are called instruments, in the
sense of instrumental variables; we avoid this term here, since it could be
confused with the financial instruments, i.e., the assets themselves. Fitting
the model by least squares, the factors are
${\bf f}_t=({\bf \Gamma}^\T\G{\bf \Gamma})^{-1}{\bf \Gamma}^\T\P^\T{\bf r}_t$
for a given ${\bf \Gamma}$, and the coefficients maximise
\begin{equation}
\label{eq:rayleigh}
\Tr\Bigl[\bigl({\bf \Gamma}^\T\G\,{\bf \Gamma}\bigr)^{-1}\,
{\bf \Gamma}^\T\H\,{\bf \Gamma}\Bigr] , \qquad
\G=\P^\T\P , \quad \H=\P^\T\C\,\P .
\end{equation}
The maximum is reached when the columns of ${\bf \Gamma}$ span the $K'$ leading
solutions of the generalised eigenvalue problem $\H{\bf a}=\mu\,\G{\bf a}$ of
Section~\ref{sec:theory}. The estimated loadings are therefore the first $K'$
constrained eigenvectors, i.e., the first $K'$ columns of $\Z$ in
Eq.~(\ref{eq:Z}), and the variances of the factors are the constrained
eigenvalues $\mu_k$.

Two known methods are recovered in this way. The first one is the
\textit{projected principal component analysis} of \citet{Fan16}, to which the principal
component analysis under constraint of Section~\ref{sec:theory} is equivalent.
Let ${\bf \Pi}=\P\G^{-1}\P^\T=\Z\Z^\T$ be the projector onto the subspace
spanned by the basis portfolios. A vector ${\bf w}$ of this subspace obeys
${\bf w}={\bf \Pi}{\bf w}$, so that its variance is ${\bf w}^\T\C\,{\bf w}=
{\bf w}^\T{\bf \Pi}\,\C\,{\bf \Pi}\,{\bf w}$: maximising the variance under
the constraint that ${\bf w}$ belongs to the subspace is the same as
diagonalising ${\bf \Pi}\,\C\,{\bf \Pi}$, the correlation matrix of the
projected returns ${\bf \Pi}{\bf r}_t$. Indeed,
\begin{equation}
\label{eq:equivalence}
{\bf \Pi}\,\C\,{\bf \Pi}\,\Z = \Z\,\M ,
\end{equation}
so that the constrained eigenvectors and eigenvalues are the eigenvectors and
the non-zero eigenvalues of ${\bf \Pi}\,\C\,{\bf \Pi}$. The two formulations
describe the same computation: the constraint is imposed on the eigenvectors
in one case, and on the data in the other. Second, $\H$ is the covariance matrix of
the returns $\P^\T{\bf r}_t$ of the $K$ basis portfolios, which
\citet{Kelly19} call managed portfolios. In their instrumented principal
component analysis (IPCA), the weights of each basis portfolio are the values
of an observed characteristic of the stocks, such as the book-to-market ratio,
which vary in time, and the objective is a sum over time of terms of the form
(\ref{eq:rayleigh}); when the characteristics are constant, they note that the
solution is given by the principal components of the managed portfolios, which
is the generalised eigenvalue problem above. The constrained eigenvectors of
Section~\ref{sec:theory} are thus the loadings of a static IPCA whose basis
portfolios are the maximum-variance portfolios of the groups, with all the $K$
factors retained.

\subsection{What is specific to the MaxVar estimator}

\paragraph{The weights inside each group are estimated} In the projected and
instrumented analyses, the weights of the basis portfolios are data: they are
the observed values of a characteristic for each asset, and nothing is
estimated at this stage. If the only available information is the membership
of the assets in groups, the corresponding basis portfolio of a group is its
indicator variable, i.e., equal weights on the assets of the group and zero
elsewhere. All the assets of a group then have the same loadings, and the
systematic part of the matrix is constant inside each block: a contract that
is weakly correlated with its class, such as the BTP among the bonds in
Section~\ref{sec:backtest}, is treated like the others. The MaxVar estimator
uses instead the maximum-variance portfolio of each group as a basis portfolio
(Step~1). The external information decides which assets enter each basis
portfolio, and the data decide their weights, which are the components of the
leading eigenvector of the group. The loadings inside a group are therefore
different from one asset to another (Eq.~(\ref{eq:disjoint})), and the
heterogeneity of the exposures is preserved.
This is the step that gives its name to the method, and it has no equivalent in
the two analyses above.

\paragraph{A complete correlation matrix} The models above describe the
systematic part of the returns and are used to estimate factors and risk
premia. Our purpose is to obtain a correlation matrix that can be inverted with
few observations. Steps~3 and~4 complete the factor part with a residual
variance $\delta$ that preserves the trace, and
Proposition~\ref{prop:prop} guarantees that the result is positive definite
even when the empirical matrix is singular. The matrix $\F=\delta\,\I_n+
\Z(\M-\delta\,\I_K)\Z^\T$ has the form of the covariance matrix of
probabilistic principal component analysis \citep{Tipping99}, with the
constrained eigenvectors in place of the unconstrained ones.

\paragraph{Risk rather than expected returns} The asset-pricing literature
uses the characteristics to explain average returns on large panels of stocks.
Here the groups are used to reduce the measurement noise of the correlations
when the number of observations is comparable to the number of assets, and the
estimator is assessed by the risk-adjusted performance of a portfolio that
involves the inverse matrix.

\subsection{Characteristics and rank-based loadings also derived from constrained diagonalisation}
\label{app:rank}

For single stocks, the groups are replaced by financial characteristics such as
the book value, the capitalisation or the momentum, and the maximum-variance
portfolio of a characteristic is defined as the market-neutral portfolio that
maximises its variance, for a given norm of the weights \citep{Valeyre19}.

The loadings are measured, once again, by a diagonalisation under constraint.
For each style, the stocks are sorted according to the characteristic and
grouped into ten quantile portfolios of equal size, inside which each stock is
weighted by the inverse of its volatility. The correlation matrix is then
diagonalised in the subspace spanned by these ten portfolios, exactly as in
Section~\ref{sec:theory}, with the quantile portfolios in the role of the
group portfolios. The loadings are the weights of the first market-neutral
constrained eigenvector, which is a linear combination of the ten quantile
portfolios: they are not postulated, but read from the direction of maximum
variance. \citet{Valeyre19} measured them in this way as a function of the
rank for 14 different styles (Fig.~\ref{fig:BM-ERW(6,9)ALL}), and found them
to be linear in the rank. For
the 500 largest US stocks, they found that the sensitivity of
stock $i$ to the corresponding factor is well described by $\Sigma_i\,B(q_i)$,
where $\Sigma_i$ is the volatility of the stock, $q_i$ its rank among the $N$
stocks according to the characteristic, and
\begin{equation}
\label{eq:rank}
B(q) = \frac{q-1}{N-1} - \frac{1}{2} .
\end{equation}
The sensitivities are thus linear in the rank, i.e., uniformly distributed
among the stocks, and the same function $B$ describes nine of the fourteen
factors that were studied. The weights of the maximum-variance portfolio are
then proportional to the centred ranks.

\begin{figure}[!htbp]
	\begin{center}
		\includegraphics[width=75mm]{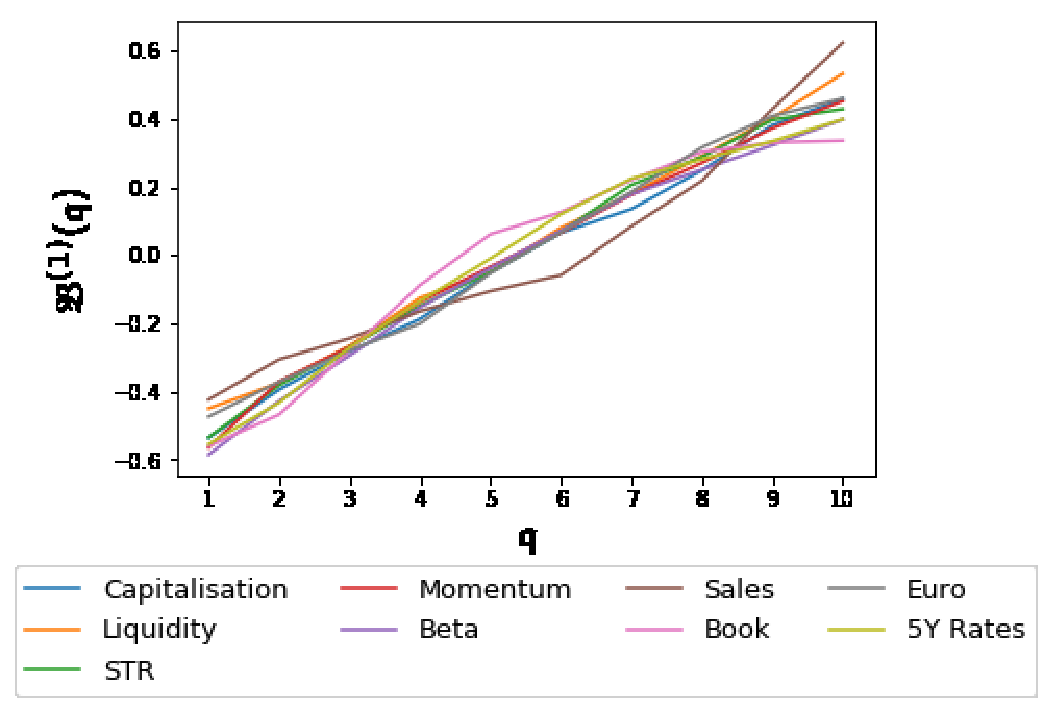}
		\includegraphics[width=65mm]{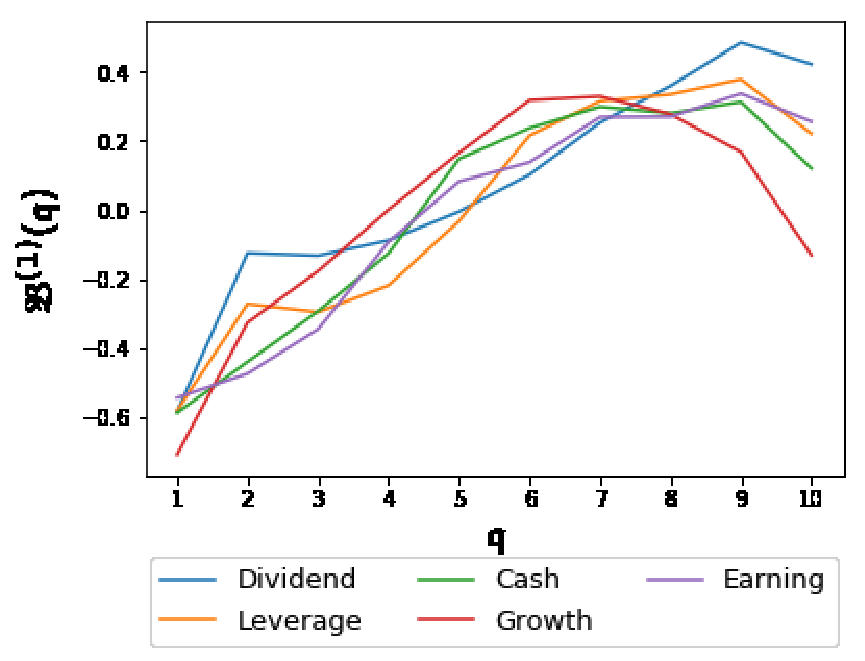}
	\end{center}
	\caption{Function $\boldsymbol{\mathfrak{B}}\left(\mathfrak{q}\right)$
measured by \citet{Valeyre19}, which gives the sensitivity of a stock to a
factor as a function of its rank $\mathfrak{q}$. Top: the nine factors for
which the function is linear. Bottom: the remaining five factors, for which the
function has a different shape, which could be explained either by noise or by
the convergence towards another risk factor; according to \citet{Valeyre19},
these five signals capture risk factors that are too weak for
$\boldsymbol{\mathfrak{B}}\left(\mathfrak{q}\right)$ to be measured properly.
The normalisation is the same for all the factors.}
	\label{fig:BM-ERW(6,9)ALL}
\end{figure}

This universality is intriguing. One could have expected the sensitivities to
follow the distribution of the characteristic itself, which is close to a
Gaussian for most financial ratios, and hence to be larger for the stocks with
extreme values. They are instead uniformly distributed, whatever the
characteristic. The thesis in which this result was reported has remained
little cited, and, to our knowledge, the result has been neither taken up nor
tested by other authors.

A possible interpretation is the following. Investors who follow a style do
not use the value of a characteristic but the ordering of the stocks. A typical
style portfolio, in the academic literature since \citet{Fama93} as well as in
the industry, is long the fraction $\theta$ of the stocks with the highest
values of the characteristic and short the fraction $\theta$ with the lowest
ones. If the common factor associated with a characteristic is driven by the
flows into and out of such portfolios, as in the style-investing model of
\citet{Barberis03}, the sensitivity of a stock is proportional to the net
number of portfolios that hold it. For a stock of normalised rank
$u=(q-1)/(N-1)$ and for cut-offs $\theta$ uniformly distributed between $0$ and
$1/2$, i.e., without any preferred value,
\begin{equation}
\label{eq:quantiles}
\int_0^{1/2}\Bigl[{\bf 1}_{\{u>1-\theta\}}-{\bf 1}_{\{u<\theta\}}\Bigr]\,
\mathrm{d}\theta = u-\frac{1}{2} ,
\end{equation}
which is exactly the law (\ref{eq:rank}). In this reading, the linearity in the
rank reflects the absence of a preferred quantile among investors, and its
universality follows from the fact that the argument involves only the sorting
procedure, and neither the nature of the characteristic nor its distribution.
This interpretation is a conjecture that we have not tested, and two
qualifications are in order. First, the absence of a preferred quantile must
hold for the aggregate weight placed on each cut-off, and not only for the
number of investors: if those who trade the extreme deciles manage more
capital, or take larger positions per stock because their portfolios contain
fewer names, the exposure is no longer linear in the rank. Second, the argument
gives the aggregate holdings as a function of the rank; to obtain the same
shape for the sensitivities of the returns, the flows into a style must also
move the prices in proportion to these holdings, once the volatility is taken
into account. The interpretation has a testable prediction: when a specific
cut-off dominates, the exposure should change visibly around it, for instance
around the 10th and 90th percentiles for portfolios built on deciles. In
Fig.~\ref{fig:BM-ERW(6,9)ALL}, the departures of the five characteristics that
do not follow the common shape are located in the extreme deciles: the first
decile lies well below the line, and the curves flatten or decrease over the
last two or three deciles, whereas they are close to linear in between. This is
in line with the prediction, but two other explanations cannot be excluded. The
extreme deciles of these characteristics contain specific populations, such as
stocks that pay no dividend or have negative earnings, and \citet{Valeyre19}
attribute the departures to factors that are too weak to be measured
accurately. A measurement with more than ten quantiles would help to
discriminate between these explanations.

Weights that are linear in the rank are commonly used to build factor
portfolios \citep{Asness13,Frazzini14}, and \citet{Kelly19} transform each
characteristic into exactly the centred rank (\ref{eq:rank}) before using it
as the weights of a managed portfolio. In both cases the motivation is the robustness to outliers.
The result of \citet{Valeyre19} provides an empirical basis for this choice:
the rank-linear form is not only convenient, it is the form of the loadings
that is measured when the variance captured by a characteristic is maximised,
and it is common to most characteristics. To our knowledge, this property has
not been reported as an empirical result. The closest measurements are the
characteristic-beta functions estimated semiparametrically by \citet{Connor07}
and \citet{Connor12}. There, each characteristic has its own curve, which is
expressed as a function of the normalised characteristic itself and not of its
rank: on that scale, the curves for size and value are concave, and those for
momentum and volatility are closer to linear. The two descriptions are
compatible, since a curve can be nonlinear in the characteristic and linear in
its rank. What is specific to \citet{Valeyre19} is the change of variable:
expressed as a function of the rank, and after scaling by the volatility, a
single linear shape fits nine of the fourteen characteristics. With such rank-based
portfolios as basis portfolios, the constrained diagonalisation of
Section~\ref{sec:theory} is the static counterpart of the IPCA, obtained
independently and from the point of view of risk rather than of expected
returns.

\section{Relation to other estimators that constrain the eigenvectors}
\label{app:shrinkhc}

\subsection{Estimators based on predefined groups}
\label{app:groups}

\paragraph{Group averages} The first estimator that uses a classification of
the assets is the mean model of \citet{EltonGruber73}: the correlation of two
assets is replaced by the average correlation inside their group, or between
their two groups. The matrix is constant inside each block, so that all the
assets of a group are treated alike. The MaxVar estimator keeps instead a
loading for each asset.

\paragraph{Hierarchical principal component analysis} In the HPCA of
\citet{Avellaneda20}, the first principal component $F_k$ of each group is
used as a factor, and the residuals of assets that belong to different groups
are assumed to be uncorrelated. The correlation of asset $i$ in group $k$ and
asset $j$ in group $l\ne k$ is then $\beta_i\,\beta_j\,\rho_{kl}$, where
$\beta_i$ is the correlation of asset $i$ with the factor of its group and
$\rho_{kl}$ the correlation of the factors $F_k$ and $F_l$, whereas the
empirical correlations are kept inside each group. In the notations of
Section~\ref{sec:theory}, $\beta_i=\sqrt{\lambda_k^{(0)}}\,p_{k,i}$ and
$\rho_{kl}=\H_{kl}/\sqrt{\lambda_k^{(0)}\lambda_l^{(0)}}$, so that
\begin{equation}
\beta_i\,\beta_j\,\rho_{kl} = \H_{kl}\, p_{k,i}\, p_{l,j} ,
\end{equation}
which is the second line of Eq.~(\ref{eq:disjoint}). For disjoint groups, the
MaxVar matrix $\F$ and the HPCA matrix therefore coincide for all the pairs of
assets that belong to different groups. The idea of using the maximum-variance
portfolio of each predefined group to describe the correlations between groups
is thus common to the two approaches. They were developed independently and
almost simultaneously: the maximum-variance portfolios of predefined
characteristics and sectors, and the diagonalisation of the correlation matrix
in the subspace that they span, appear in a chapter dated January 2019 of a
thesis defended in May 2019 \citep{Valeyre19}, and the HPCA in a preprint of
October 2019 \citep{Avellaneda20}. One difference of detail is that the
portfolio of each theme is market neutral in \citet{Valeyre19}: it is the
second eigenvector of the theme, after that of the market, whereas the HPCA
and the present paper use the first one.

The two estimators differ inside the groups. The HPCA keeps the empirical
correlations, and with them both the genuine structure and the measurement
noise of each block. In particular, the HPCA matrix inherits the null space of
the empirical blocks: it is singular as soon as a group contains more assets
than there are observations, which is the case for the short estimation window
of Section~\ref{sec:backtest}. The MaxVar estimator replaces the correlations
inside a group by those of a one-factor model with a common residual variance
(first line of Eq.~(\ref{eq:disjoint})). It has far fewer parameters and is
always invertible (Proposition~\ref{prop:prop}), at the price of the bias
discussed in Section~\ref{sec:bias}. In addition, the constrained
diagonalisation of Section~\ref{sec:theory} handles groups that overlap, such
as asset classes and regions used together.

\paragraph{Factor models with a cleaned residual} \citet{DeNard21} follow a
third route: the factor part is estimated by regression on observed or
statistical factors, and the covariance matrix of the residuals is neither
kept as it is nor replaced by a diagonal matrix, but cleaned by nonlinear
shrinkage. The three approaches can thus be ordered by the way in which they
treat what the group factors do not explain: the HPCA keeps the empirical
residual correlations inside the groups, \citet{DeNard21} shrink them, and the
MaxVar estimator replaces them by a flat spectrum. The more structure is
imposed, the smaller the variance of the estimator and the larger its possible
bias, so that the best choice depends on the number of observations per asset.
Applying a nonlinear shrinkage to the residual of the MaxVar decomposition, in
the orthogonal complement of the group portfolios, would be the natural
intermediate between the MaxVar estimator and the HPCA.

\subsection{Eigenvector shrinkage}

The estimators of the first family of the Introduction leave the empirical
eigenvectors untouched. \citet{Goldberg22} showed that, when the number of
assets is large compared with the number of observations, the leading empirical
eigenvector is systematically biased, and that this bias, rather than the error
on the eigenvalues, is what degrades minimum-variance portfolios. They correct
it by shrinking the leading eigenvector towards the vector whose components are
all equal, with an intensity estimated from the data; \citet{Goldberg23}
interpret this correction as a James--Stein estimator for eigenvectors.
\citet{Gurdogan22} extend it to several anchor directions, for instance the
indicator vectors of sectors, and \citet{Goldberg24} choose the anchor
directions from the constraints of the portfolio that is optimised.

The MaxVar estimator shares the starting point of these works: prior
directions are used to improve the eigenvectors, which no rotationally
invariant estimator can do. It differs from them in three respects. First, the
constraint is hard instead of soft: the leading eigenvectors are projected onto
the subspace of the group portfolios, which corresponds to a shrinkage of
intensity one, whereas the James--Stein approach interpolates between the
empirical eigenvector and its anchor with an intensity that has optimality
properties in a one-factor model. Using the MaxVar matrix as a shrinkage
target, as suggested in Section~\ref{sec:conclusion}, would be the counterpart
of this interpolation. Second, the prior directions are not fixed vectors: only
the composition of the groups is given, and the weights inside each group are
estimated (\ref{app:ipca}), whereas an indicator vector gives the same weight
to all the assets of a sector. Third, $K$ eigenvectors are constrained
simultaneously, one per group, instead of the leading one.

\subsection{Hierarchical clustering}

A second way to constrain the eigenvectors is to infer a hierarchy of clusters
from the data. Since \citet{Mantegna99}, the dendrogram obtained by
hierarchical clustering of the correlation matrix has been used to filter it:
in the hierarchically nested factor model of \citet{Tumminello07a}, each node
of the dendrogram is a factor, and the filtered correlation of two assets is
the one attached to the node at which they merge
\citep{Tumminello07b,Pantaleo11}. \citet{Bongiorno21} average this filtered
matrix over bootstrap replicas of the data, which makes it more flexible, and
\citet{Bongiorno22} use such estimators dynamically. The same
hierarchy can be used directly at the level of the allocation: the
hierarchical risk parity of \citet{LopezdePrado16} allocates the risk along
the dendrogram without inverting the matrix, and \citet{Cotton24} shows that
this allocation and the minimum-variance portfolio are two limits of a single
family based on Schur complements.

The comparison with the MaxVar estimator is instructive on two points.

\paragraph{Groups that are given or learnt} In the clustering approaches the
groups are an output: they adapt to the data, but they inherit the measurement
noise of the correlation matrix, which is the reason for the bootstrap
averaging of \citet{Bongiorno21}. In the MaxVar estimator the groups are an
input, which does not depend on the sample. Asset classes and regions used
together form a two-level hierarchy that is fixed in advance. The two points
of view can be combined: nothing in Section~\ref{sec:theory} requires the
groups to be external, and the clusters obtained by cutting a dendrogram at a
given level could be used as groups. One would lose the independence from the
sample and gain the ability to follow changes in the structure of the market.
We have not tested this variant.

\paragraph{Correlations inside a group} In the filtering by hierarchical
clustering, all the pairs of assets that merge at the same node receive the
same correlation, so that the filtered matrix is constant inside the blocks
defined by the dendrogram. In the MaxVar estimator, the correlations inside a
group are those of a one-factor model with asset-specific loadings
(Eq.~(\ref{eq:disjoint})): a contract that is weakly correlated with its class
keeps a small loading. In this respect, the MaxVar estimator is to the
hierarchical filtering what a one-factor model is to a constant-correlation
model.

\paragraph{Clustering in time} The works above cluster the assets.
\citet{Wang26} apply hierarchical clustering to a different object: the
covariance matrices estimated on successive sub-periods of the estimation
window. The matrices are compared through the similarity of the associated
correlation matrices and grouped accordingly, and the estimate is obtained by
aggregating them, which separates the persistent configurations of the
correlations from the transient ones. The resulting estimator is applied to
minimum-variance portfolios of exchange-traded funds. This approach uses
information that the MaxVar estimator ignores, namely the similarity between
the correlation structures observed at different dates, whereas the MaxVar
estimator uses information that it ignores, namely the membership of the assets
in groups. The two are therefore complementary rather than competing: the
matrix of each sub-period could be cleaned by the MaxVar estimator before the
aggregation. We have not tested this combination.


\end{document}